# Evaporation of droplets on strong and low-pinning surfaces and dynamics of the triple line

Edward Bormashenko, Albina Musin and Michael Zinigrad

Ariel University Center of Samaria, Applied Physics Department, Department of Chemical Engineering and Biotechnology, 40700, P.O.B. 3, Ariel, Israel

**Abstract**

Evaporation of water droplets deposited on metal and polymer substrates was studied. The evaporated droplet demonstrates different behaviors on low-pinning (polymer) and strong-pinning (metallic) surfaces. When deposited on polymer surfaces, the evaporated droplet is featured by stick-slip sliding, whereas on strong-pinning metallic surfaces it does not show such kind of motion and demonstrates the giant contact-angle hysteresis. Stick-slip motion of droplets is described satisfactorily by the Shanahan-Sefiane model relating this kind of motion to surmounting potential barriers caused by the pinning of the triple (three-phase) line. The experimentally established "stick" times coincide with the values predicted by the Shanahan-Sefiane theory. The values of potential barriers are reported. The notion of the equilibrium contact angle is refined.

## 1. Introduction

The contact angle serves as a natural characteristic of wetting at solid/liquid and liquid/liquid interfaces for two hundred years.[1-4] According to De Gennes *et al.*, solid substrates could be divided into two categories: a) high-energy surfaces, for which the chemical binding is of order of 1 eV, inherent to ionic, covalent or metallic bonds, and b) low-energy surfaces inherent, e.g., to polymers, for which the chemical binding energy is of the order of $kT$. In the first case, the solid/air interfacial tension is of order of $\gamma_{SA} \sim 500–5000$ mJ/m$^2$ and nearly all liquids including water spread on such substrates (thus the contact angle must be zero). For the low-surface-energy substrates, $\gamma_{SA} \sim 10–50$ mJ/m$^2$, and they demonstrate the partial wetting characterized by non-zero contact angles.[1-5]

It has long been known that the equilibrium (or Young) contact angle corresponding to the free-energy minimum of a droplet-substrate system is hardly observed due to the phenomenon of the contact-angle hysteresis.[6-20] Multiple minima of the free energy of a droplet deposited on a solid substrate promote multiplicity of contact angles.[14] The contact-angle-hysteresis phenomenon is related to the pinning of the triple line separating solid and liquid phases due to physical or chemical heterogeneities of the substrate. It is generally agreed that the roughness of the substrate strengthens the hysteresis significantly, whereas atomically flat surfaces demonstrate relatively low hysteresis.[6-21] We show that both the high-energy and the low-energy nano-rough surfaces a featured by very high values of the contact angle hysteresis, and neither "as placed" contact angles nor the value of the hysteresis does not characterize the surface comprehensively. At the same time, the dependence of the contact angle on the droplet radius in the course of evaporation allows distinction between the surfaces. Instead of the terms "high-" and "low-energy" we propose to describe the solid substrates in the notions of strong- and low-pinning ones. We suggest that this classification describes the behavior of evaporated droplets in a much more adequate way.

2. Experimental

Two types of surfaces were used in the experiments. Six extruded polymer substrates were used as low-energy ones, i.e. polyethylene (PE), polypropylene (PP), polysulfone (PSu), polyvinylidene fluoride (PVDF) poled and non-poled (Kynar), polyethylene terephtalate (PET). PSu and PVDF films were supplied by Westlake Plastics Co. Before the measurements all substrates were cleaned thoroughly with ethyl alcohol, rinsed with a large amount of bi-distilled water and dried.

Surfaces with high energy were: stainless steel and aluminum (Al). Stainless steel A304 and aluminum Al2024 samples were prepared using the Struers company equipment including Labotom-3 cut-off machine, LaboPress-1 mounting press and LaboPol-6 grinding and polishing machine. The process of preparation of the specimen included 3 stages. First, the specimen was sealed into a resin cylinder in the press (we used Multifast resin). After that, the specimen was grinded in 2 steps, namely, plain and fine grinding, using discs and lubricants according to instructions of the Struers company. The final stage was polishing; all specimens were polished using MD-Chem disc (metal backed porous synthetic polishing cloths) with OP-S abrasive (colloidal silica suspensions with grain size of 0.04 µm,) with addition of DP-Blue lubricant (alcohol-based suspension). Finally the specimens were rinsed thoroughly with ethanol, distilled water and dried.

Roughness of the surfaces ($R_a$) measured according to AFM Nano Scope 5 equipped with Gwyddion 2.19 software is supplied in the Table 1. It could be seen that all studied substrates were nano-rough.

**3. Results and discussion**

*3.1. Evaporation of droplets on metallic surfaces*

Let us start from high-energy (metallic) surfaces. A 10 µl droplet was placed on thoroughly cleaned surfaces and evaporated. Large "as placed" angles (in the notions proposed in Ref. 22) for steel, as high as 70º, attract attention. Large contact angles observed on metallic surfaces were reported also by other groups. Abdelsalam *et al* reported a value of 70º as a contact angle for gold.[23] Iveson *et al* observed ore a contact angle as high as 74º on iron.[24, 25] Contact angles as high as 70º were observed on Ni.[26] Wang reported contact angles in the range of 68–74º on the polished stainless steel.[27, 28] Of course, the oxide film covering the metal surfaces is also involved in the formation of "as placed" angles, however the presence of this film does not convert the surface in the "low-energy" one, it remains still a high-energy surface. All these experimental results support our observations, but definitely contradict the idea that high-energy surfaces have to be completely wetted. Bewig and Zisman supposed that high contact angles observed on metallic surfaces are due to organic contaminants, and "in order to rid these metal surfaces of adsorbed hydrophobic contaminants, it is necessary to heat them to white-hot temperatures in flowing streams of high purity gases".[29]

A diversity of factors besides organic contamination could be responsible for high contact "as placed" angles observed on metallic surfaces. It looks reasonable to relate the high values of "as placed" contact angles to the micro-roughness of the high-energy surfaces. However, it could be seen that roughness is not responsible for this effect. Indeed, roughness exerts an impact on the wettability of surfaces according to two main scenarios, i.e. following the Cassie or Wenzel wettability models.[3, 30, 31] According to the Cassie model, air is trapped below a droplet, forming "air pockets".

The Cassie wetting regime is featured by low hysteresis of the apparent contact angle.[32] This obviously contradicts to our observations: the contact angle hysteresis as high as 40–50º was registered on steel and Al substrates (see Fig. 1a and 1b). According to the Wenzel model, the roughness increases the wetted area of a solid, which also geometrically modifies hydrophobicity; thus inherently hydrophobic surfaces become more hydrophobic, and inherently hydrophilic ones become more hydrophilic. The Wenzel model predicts that roughness will strengthen an inherent hydrophilicity of high-energy surfaces, thus it could not be invoked for explanation of high "as placed" angles registered on metallic substrates.

Now let us discuss the experimental data describing the droplet evaporation on metal surfaces. At the first stage of evaporation a droplet remains pinned to the substrate and the contact angle is decreased from about 70º to 20º, demonstrating the giant hysteresis of the contact angle. The further evaporation is followed by de-pinning of the three-phase line. The radius of the contact area decreases, and the contact angle continues to fall to a values about 5° as depicted in Fig. 1a and 1b.

Residual organic contamination of metallic surface, perhaps, explains high values of "as placed" angles but it definitely does not explain the giant contact angle hysteresis observed on polished and degreased metals. We suggest that the true physical reason explaining both high values of contact angles and giant hysteresis registered on high-energy surfaces is the effect of pinning of the triple (three-phase) line discussed in detail by Yaminsky.[21] Yaminsky argued that the triple line will be pinned to the surface even when substrates are atomically flat and uniform, and interaction similar to dry friction occurs at the three-phase line. Zero contact angle, which is thermodynamically favorable, remains unattainable due to potential barrier produced by the pinning of the triple line to the substrate. The crucial impact exerted by the triple line on the wetting phenomena was discussed in the series of recent papers.[33-35]

*3.2. Evaporation of droplets on polymer surfaces*

Figure 1c and d depicts changes in the contact angle and the contact radius of a water droplet with the same volume of 10 µl during evaporation on low-energy polymer (PSu and PP) surfaces. At the first stage a triple line is pinned, as on high-energy substrates, and the contact angle decreases from about 80 to 65°. But this stage is followed with a stick-slip motion of a triple line when the contact radius jumps to smaller values, and the contact angle may increase again to some extent.

Actually high-energy (metallic) surfaces demonstrate the "as placed" contact angles close to values inherent to low-energy (polymer) substrates. The reasonable question is: what is the actual difference in the wetting behavior of low- and high-energy surfaces? In order to answer this question we have to compare graphs describing the dependence of the contact angle on the radius of contact area (see Fig. 2). Two distinct portions of the curve could be recognized for high-energy substrates: 1) evaporation of a droplet when the three-phase line is pinned (the radius of the contact area is constant) accompanied by the decrease in the contact angle; 2) fast decrease of a contact radius accompanied by the slower decrease in the contact angle.

The same areas are also seen at the curves obtained with polymeric substrates. However, the low-energy surfaces demonstrate somewhat more complicated behavior. The graphs for low-energy substrates include a step with a pinned triple line as well as on high-energy ones, but it is followed with a stick-slip behavior when a contact radius decreases, steadily or with jumps, and the contact angle oscillates around some value. These oscillations may be more or less pronounced. This stage was also observed by other investigators.[36-40] Erbil *et al.* suggested that the average value of the contact angle at this stage corresponds to a true value of receding angle on smooth surfaces.[37] The stick-slip motion of evaporated drops occurring under constant contact angle was observed for various polymers, including polyethylene, polypropylene, polyethylene terephthalate, and polysulfone.[16] This kind of motion could be related to the weak interaction of a droplet with a polymer substrate, resulting in the weak pinning of a triple line, promoting the non-hysteresis sliding of a droplet. These two types of contact line movement are depicted schematically in Fig. 3.

Thus, we suppose that a new classification of surfaces should be introduced according to the dynamics of a triple line under drop's evaporation. It is reasonable to sort solid surfaces as strong-pinning (metal) and low-pinning (polymer) ones. It also should be mentioned that the notion of receding contact angle becomes irrelevant for characterization of both nano-rough metal and polymer substrates.[16] The receding contact angle defined as the minimal possible contact angle for the certain solid/liquid pair turns out to be zero (see also the discussion in Ref. 16).

*3.3. Analysis of stick-slip motion: estimation of pinning time*

It is seen from the graphs of the contact angle and the contact radius vs. evaporation time (Fig. 1 c, d) that the stick time varies (especially the time till the first jump of the contact line) depending on the substrate. To compare stick times on various substrates we use the model of the evaporating droplet proposed by M. Shanahan *et al.*[36, 40]. A droplet in the absence of gravitational flattening is represented by a spherical cap with the volume $V$

$$V = \frac{\pi R^3}{3} \frac{(1-\cos\theta)^2 (2+\cos\theta)}{\sin^3\theta}, \tag{1}$$

where $R$ is contact radius and $\theta$ is contact angle (see Fig. 4). On the stage when a contact line is pinned and the contact radius $R$ is constant, a contact angle changes from $\theta_0$ corresponding to the initial equilibrium state to a threshold value $\theta_t$ at which a movement of triple line begins. The volume evaporation rate may be calculated as:

$$\frac{dV}{dt} = \frac{dV}{d\theta}\frac{d\theta}{dt} = \frac{\pi R^3}{(1+\cos\theta)^2}\frac{d\theta}{dt}. \quad (2)$$

After integrating Eq. (2) between $\theta = \theta_0$ and $\theta = \theta_t$ the stick time is given by

$$t_{st} = -\frac{\pi R^3 \delta\theta}{(1+\cos\theta_0)^2 \, dV/dt} \quad (3)$$

where $\delta\theta = \theta_0 - \theta_t$. The volume evaporation rate $dV/dt$ is negative and may be calculated from the experiments, as well as $\theta_0$, $\theta_t$, and $\delta\theta$. Table 2 presents times of pinning (stick times) till the first jump of a triple line for 6 various polymer substrates. Two values are included – calculated according to Eq. (3) and measured directly on the graph. Taking into account the variability of the evaporation data measured on the same substrate in different points, the matching of calculated and measured values is quite convincing with the only exception of the poled PVDF. The calculated stick time for poled PVDF is nearly twice larger that the experimental value. This discrepancy may be due to the very unusual structure of the poled PVDF resulting in its ferroelectric properties, i.e. spontaneous polarization.[41, 42]

*3.4. Changes in surface free energy during evaporation*

According to model of evaporation based on considerations of the excess free energy, a droplet placed on a substrate is initially in the equilibrium state with the contact radius $R_0$ and the contact angle $\theta_0$. The corresponding Gibbs free energy $G$ is given by[36, 40]

$$G = \gamma A + \pi R^2 (\gamma_{SL} - \gamma_{SV}), \quad (4)$$

where $\gamma$, $\gamma_{SL}$, $\gamma_{SV}$ are the surface tensions at the liquid-vapor, solid-liquid and solid-vapor interfaces, respectively.

The surface of the liquid-vapor interface $A$ may be calculated as a spherical cap surface

$$A = \frac{2\pi R^2}{(1+\cos\theta)}.  \quad (5)$$

Using the Young equation $\gamma\cos\theta_0 = \gamma_{SV} - \gamma_{SL}$ and Eq. (5), $G$ in Eq. (4) may be written as

$$G = \gamma\pi R^2 \left[\frac{2}{(1+\cos\theta)} - \cos\theta_0\right]. \quad (6)$$

During evaporation, a droplet may be in a state with a larger contact radius than the equilibrium one (for the same volume of the droplet), $R = R_0 + \delta R$, and with the smaller contact angle, $\theta = \theta_0 - \delta\theta$. The corresponding *excess* of the free energy is $\delta G = G(R) - G(R_0)$. This excess free energy per unit length of a triple line was evaluated by M. Shanahan[40] as:

$$\delta\tilde{G} = \frac{\delta G}{2\pi R} = \frac{\gamma R(\delta\theta)^2}{2(2+\cos\theta_0)}. \quad (7)$$

We supposed that the initial state with "as placed" contact angle $\theta_0$ was the equilibrium state and calculated the normalized free energy according to

$$\frac{\delta\tilde{G}}{\gamma} = \frac{R(\delta\theta)^2}{2(2+\cos\theta_0)}. \quad (8)$$

Figure 5 presents the change in the normalized free energy $\delta\tilde{G}/\gamma$ during evaporation time calculated with Eq. (8) for polymer substrates – PET and PSu. Open circles correspond to values calculated with respect to the initial (as placed) contact angle $\theta_{01}$. The potential barriers inherent to the stick-slip motion of a droplet are seen clearly. In could be recognized from Fig. 5, that the free energy does not return to zero value after slip of a triple line to a new equilibrium state. But it is possible to correct this mismatch if we take the angle just after the first slip of contact line as the secondary value of the equilibrium contact angle $\theta_{02}$ as it was proposed by Shanahan and Sefiane in Ref. 40. This agrees with the situation after the slip of the contact line when the surface surrounding a droplet was wetted previously and differs from the original dry one. Solid circles depict the normalized free energy $\delta \widetilde{G}/\gamma$ calculated with Eq. (8) using two values of the equilibrium contact angle – $\theta_{01}$ before the first slip and $\theta_{02}$ after (displayed in Fig. 3). It is seen from Fig. 5 that the *excess* free energy returns to zero if we adopt $\theta_{02}$ as the new equilibrium contact angle.

It could be concluded that some accuracy is necessary when we speak about the "equilibrium contact angle". There exist two essentially different values of the equilibrium contact angle, the first of which (the Young angle) corresponds to the situation when a droplet is placed on the dry substrate, and the second corresponds to the case when a droplet is surrounded by the wet substrate. Starov and Velarde in their recent analysis stated that the vapor molecules tend to adsorb on the solid substrate, and obtained the value of the equilibrium contact angle in the situation when a droplet contacts with a solid coated with the absorbed layer of liquid.[43] The equilibrium contact angle predicted by the thermodynamic analysis reported in Ref. 43 is different from the Young one. Perhaps, it could be identified with the contact angle $\theta_{02}$, introduced on Ref. 40 and reported in our paper.

*3.5. Calculation of the energy barriers*

The graph of excess free energy (Fig. 5) presents several energy barriers that the evaporating droplet overcomes during stick-slip motion. This curve is a "negative" relative to the graph of contact angle vs. time. Peaks of excess free energy correspond to minimal angle just before de-pinning, the first peak being more pronounced corresponding to the greatest decrease in the contact angle from its initial value $\theta_0$. But there is some critical value of contact radius after which the dynamics of evaporation changes. Unlike the preceding cycles, the contact angle does not oscillate but decreases steadily (starting from the value labeled $\theta_f$, see Fig. 2-3) and the contact radius decreases slowly or even stays constant until the disappearance of the droplet. This was explained[36, 40] by insufficiency of capillary excess free energy to overcome the barrier $U$. If we find this critical radius $R_c$ and calculate the corresponding free energy $G_c$ with Eq. (6) it may be used to evaluate potential energy barrier $U$:

$$G_c = \gamma \pi R_c^2 \left[ \frac{2}{(1+\cos\theta_f)} - \cos\theta_0 \right] < 2\pi R_c U, \quad U \sim \gamma R_c \left[ \frac{1}{(1+\cos\theta_f)} - \frac{1}{2}\cos\theta_0 \right] \quad (9)$$

Table 3 presents the values of the barrier $U$ calculated according to (9) with measured values of $R_c$, $\theta_f$, $\theta_0$ (marked in Fig. 3) for water droplets on 6 polymer substrates. For our experiments a potential energy barrier per unit length of a triple line is of order of $10^{-5}$ J/m for all kinds of polymers. The dimension of $U$ hints that it could be identified with the line tension as it was already supposed in Ref. 40. The value of $U$ established in our paper is one order of magnitude larger than that reported by Shanahan and Sefiane.[40] And it is several orders of magnitude larger than the line tension established for atomically flat surfaces.[3, 44,45] Perhaps, $U$ reported in our paper corresponds to the "effective line tension" of nano-rough surfaces where the strong pinning of the triple line by the nano-relief could be supposed.[46] The true value of the line tension remains disputable (de Gennes *et al* even spoke about the "mythos of line tension" in Ref. 3). Thus additional physical insights in the field are necessary.

It is noteworthy that for 10 μl droplets studied in our paper the value of the potential barrier $U 2\pi R$ is much larger than the energy of thermal fluctuations and much lower than the energy of evaporation of the droplet $\delta Q$, i.e.: $kT \ll U 2\pi R \ll \delta Q$. Actually this interrelation between energies makes the stick-slip motion possible. It could be easily seen that for any reasonable volume of the droplet $U 2\pi R \gg kT$ takes place. For a sake of very rough estimation it could be assumed that $\delta Q \cong \lambda (2/3)\pi R^3$, where $\lambda = 2.3 \cdot 10^9$ J/m$^3$ is the volumetric latent heat of water evaporation. It could be recognized that for the radius of $R \cong \sqrt{3U/\lambda} \cong 10^{-7}$ m the energy barrier becomes comparable to the energy of evaporation of the droplet ($U \cong 10^{-5}$ J/m). Thus, it could be expected that small droplets with $R \leq 10^{-7}$ m will evaporate without stick-slip motion.

### 4. Conclusions

We conclude that the dynamics of the triple line under evaporation of water droplets deposited on strong-pinning (metal) and low-pinning (polymer) surfaces is very different. Strong-pinning surfaces are characterized by the giant hysteresis of the triple line and do not demonstrate stick-slip motion of a droplet inherent to low pinning ones. The stick-slip motion of water droplets deposited on various polymer substrates is well described by the model proposed by Shanahan and Sefiane. The experimental values of "stick time" coincide satisfactorily with the predictions of Shanahan and Sefiane model. The stick-slip motion of a droplet is stipulated by surmounting potential barriers which are due to the pinning of the triple line. The values of these barriers are reported and discussed.

The notion of the "equilibrium contact angle" is cleared up. Two very different situations characterized by various equilibrium contact angles should be distinguished. The Young equilibrium angle is observed when a droplet is placed on a dry substrate; another value of the equilibrium angle is observed for the case, when a droplet is surrounded by the wet area, e.g., in the course of evaporation. The both situations are treated experimentally and theoretically.

**Acknowledgements**

The authors are thankful to Professor Roman Pogreb and Dr. Gene Whyman for extremely useful discussions. We thank Mrs. Natalya Litvak for help in preparing samples. The paper is dedicated to the blessed memory of Professor Yakov Yevseevitch Gegusin, prominent scientist and teacher, who attracted our attention to the physical phenomena occurring in droplets.

**Table 1.** Average roughness of the substrates.

| Substrate | $R_a$, nm |
|---|---|
| PET | 15.0 |
| PSu | 8.0 |
| PVDF nonpoled (Kynar) | 8.0 |
| PVDF poled | 15.0 |
| PP | 19.0 |
| PE | 16.0 |
| Steel | 8.0 |
| Al | 4.0 |

**Table 2.** Stick times for different polymer substrates.

| Polymer | Stick time, s | |
|---|---|---|
| | calc | exp |
| PVDF poled | 2013 | 1270 |
| PE | 1108 | 970 |
| PP | 984 | 730 |
| PVDF nonpoled (Kynar) | 868 | 850 |
| PET | 774 | 880 |
| PSu | 689 | 570 |

**Table 3.** Calculated values of potential energy barrier $U$.

| Polymer | $\theta_0$, degrees | $\theta_f$, degrees | $R_c$, mm | $U$, J/m |
|---|---|---|---|---|
| PVDF Kynar | 80 | 40 | 0.5 | $0.9 \cdot 10^{-5}$ |
| PVDF poled | 75 | 15 | 0.8 | $1.1 \cdot 10^{-5}$ |
| PE | 95 | 80 | 0.7 | $2.3 \cdot 10^{-5}$ |
| PSu | 70 | 45 | 0.8 | $1.2 \cdot 10^{-5}$ |
| PET | 65 | 30 | 0.6 | $0.7 \cdot 10^{-5}$ |
| PP | 80 | 55 | 1.5 | $3.0 \cdot 10^{-5}$ |

**Figure Captions**

**Fig. 1.** The changes in the contact angle and the contact radius of the water droplet during evaporation on steel **(a)**, Al **(b)**, PSu **(c)** and PP **(d)** surfaces.

**Fig. 2.** Dependence of the contact angle on the radius of the contact area for water droplet deposited on various substrates.

**Fig. 3.** Two types of the triple line motion during evaporation on metal (1) and nonmetal (2) surfaces.

**Fig. 4. a** Geometrical parameters of a droplet. **b** Profile of the 10 µl water droplet during evaporation on the Al substrate: the initial (left) and after 1000 s (right).

**Fig. 5.** Normalized excess free energy vs. time of evaporation on **a** PET, **b** PSu calculated with initial equilibrium contact angle $\theta_{01}$ (open circles) and with two values $\theta_{01}$ and $\theta_{02}$ (solid circles).

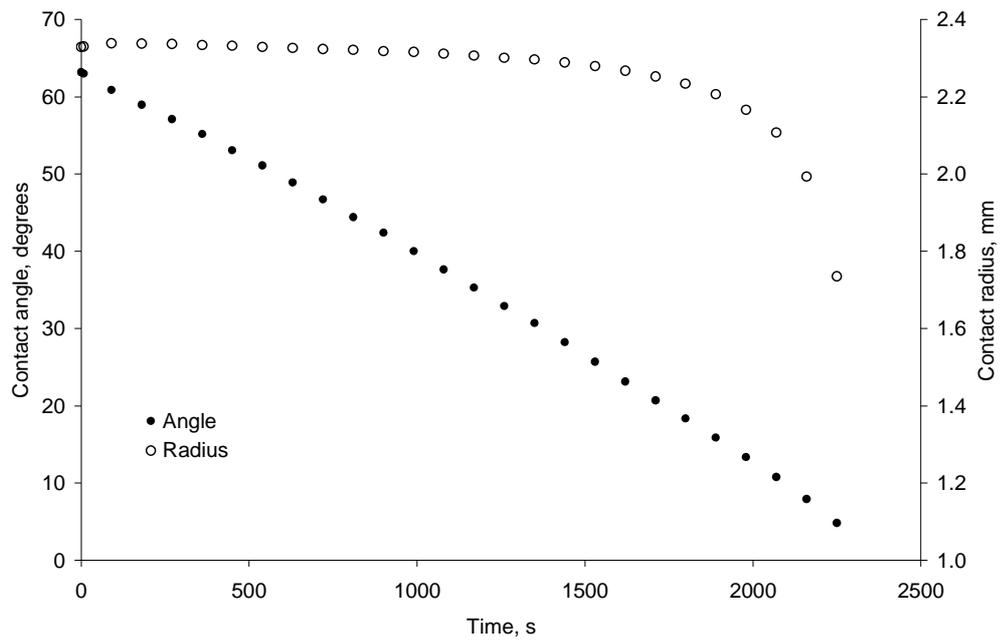

a

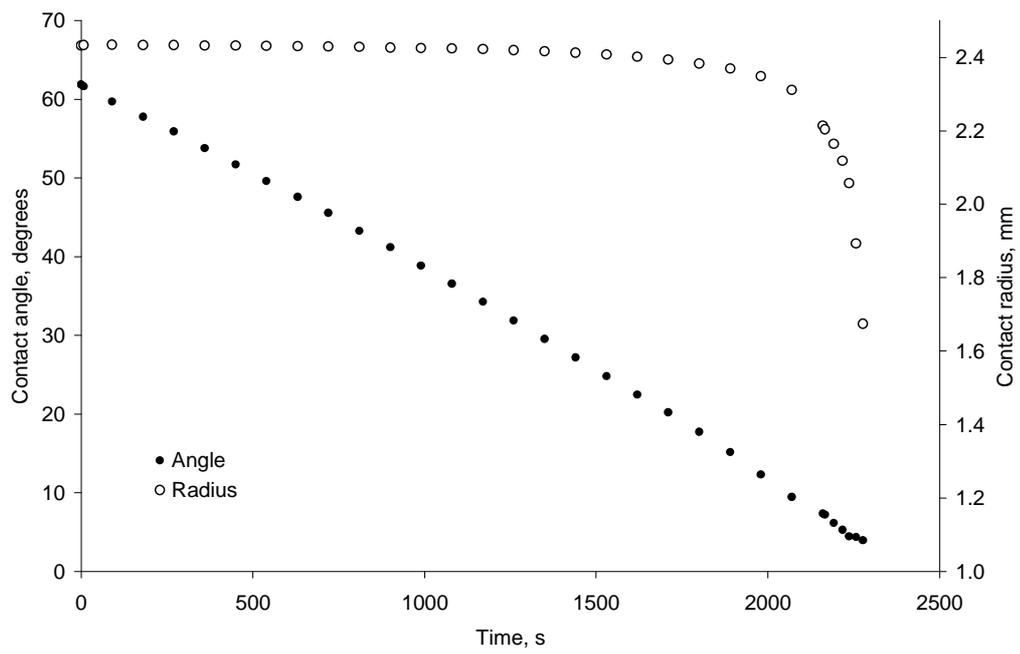

b

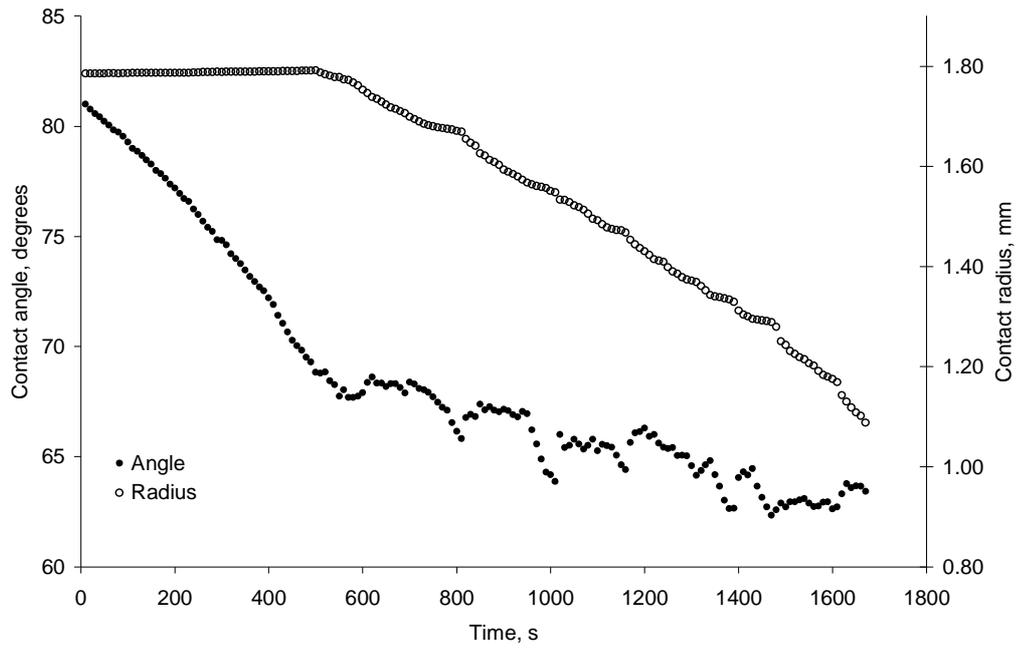

c

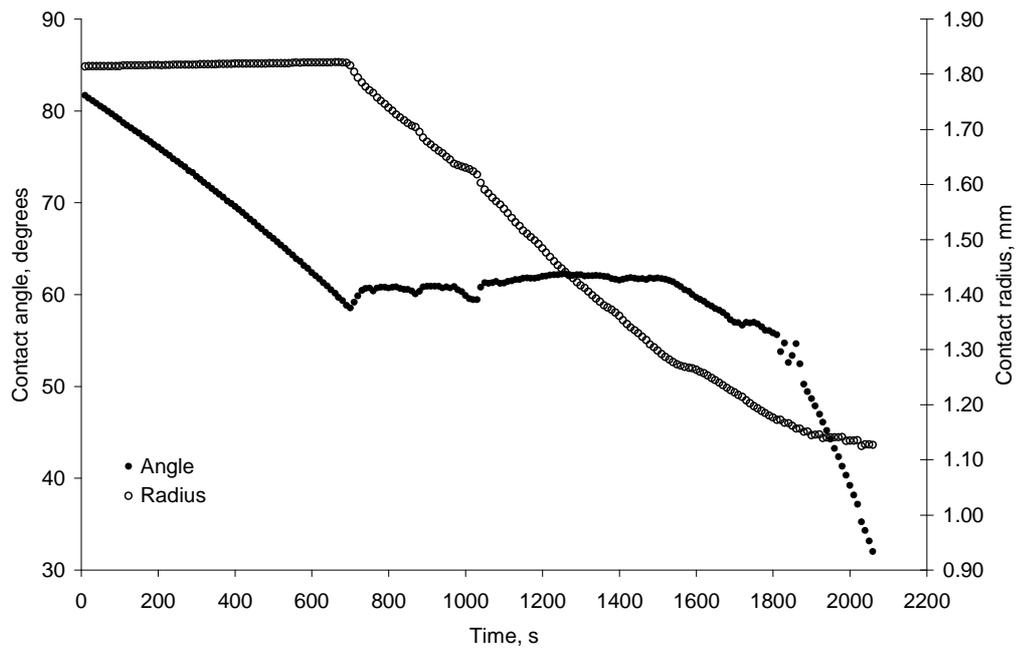

d

**Fig. 1.** The changes in the contact angle and the contact radius of the water droplet during evaporation on (**a**) steel, (**b**) Al, (**c**) PSu and (**d**) PP surfaces.

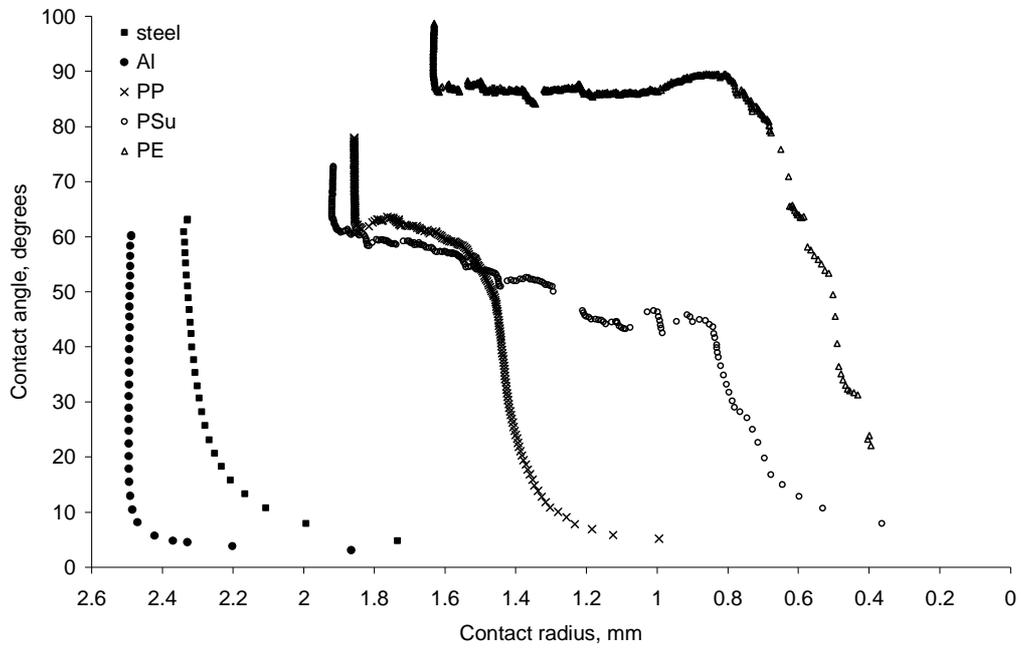

**Fig. 2.** Dependence of the contact angle on the radius of the contact area for water droplet deposited on various substrates.

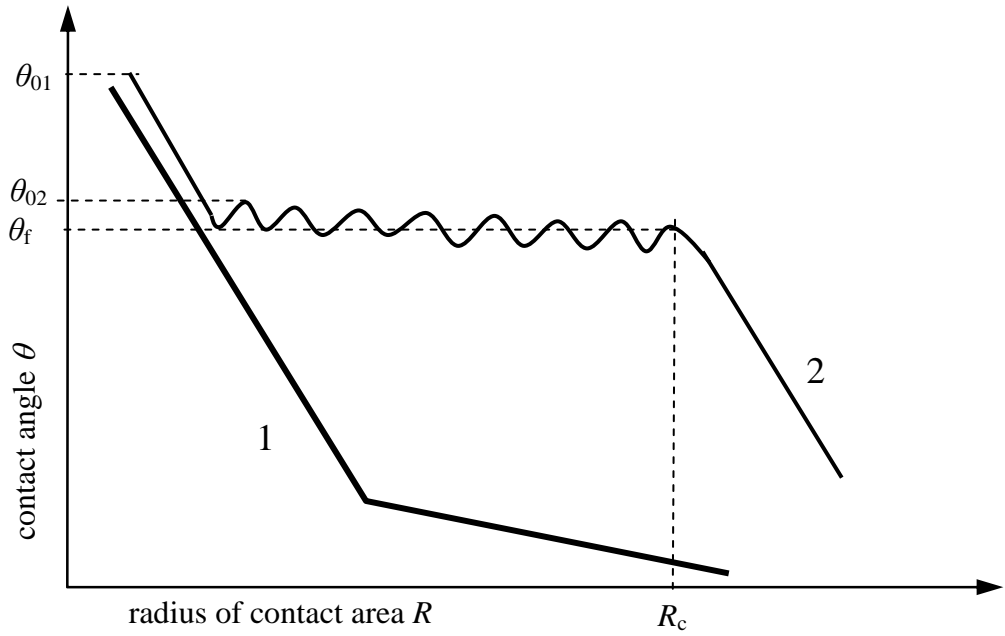

**Fig. 3.** Two types of the triple line motion during evaporation on metal (1) and nonmetal (2) surfaces.

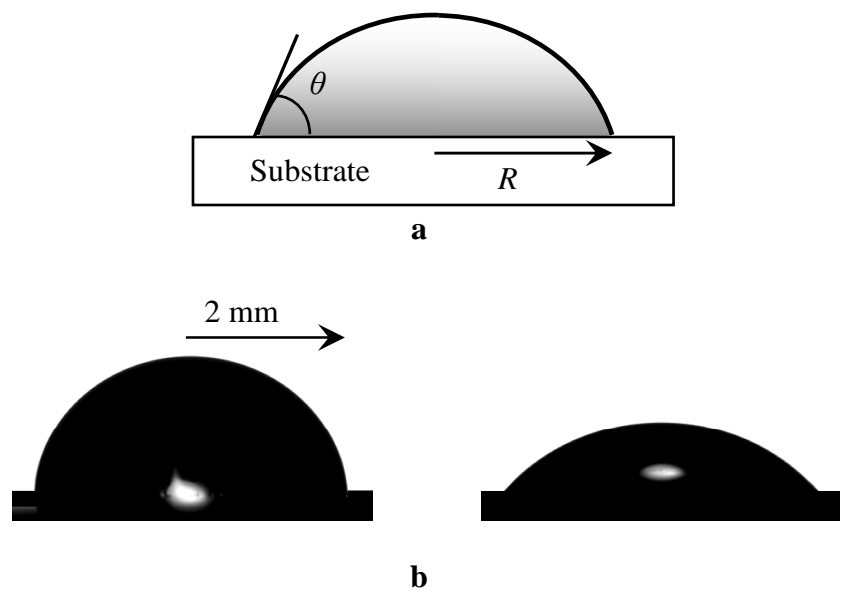

**Fig. 4. a** Geometrical parameters of a droplet. **b** Profile of the 10 µl water droplet during evaporation on Al substrate: the initial (left) and after 1000 s (right).

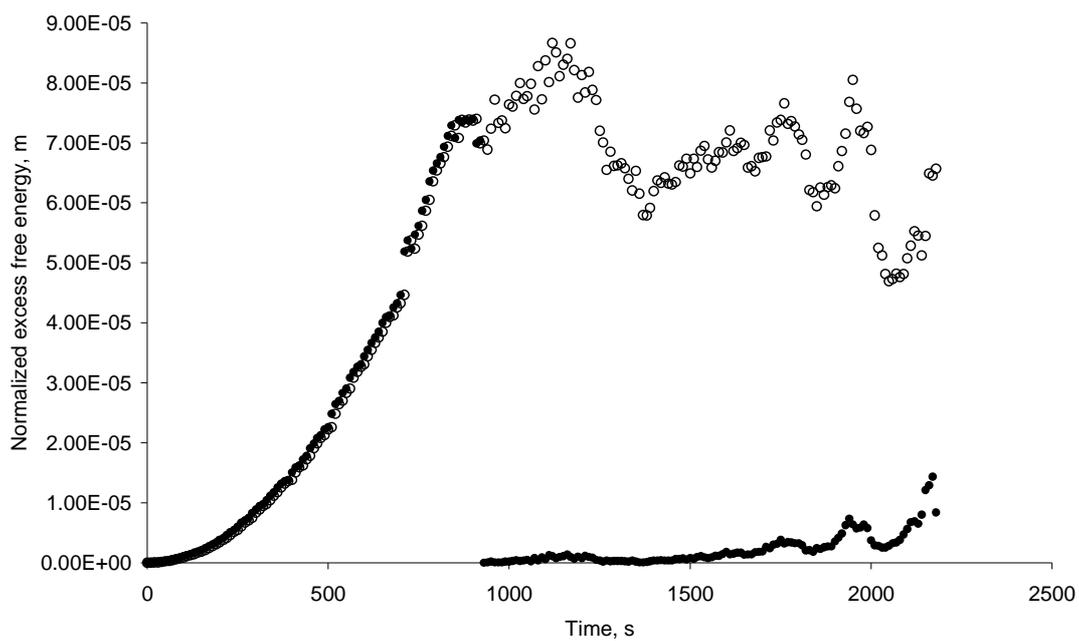

**a**

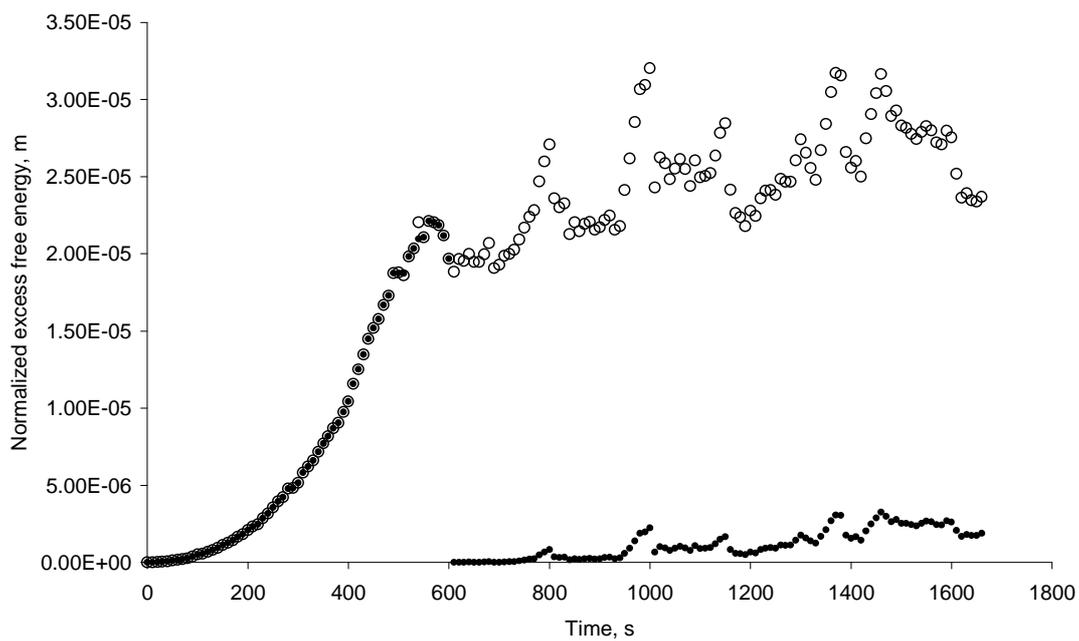

**b**

**Fig. 5.** Normalized excess free energy vs. time of evaporation on **a** PET, **b** PSu calculated with initial equilibrium contact angle $\theta_{01}$ (open circles) and with two values $\theta_{01}$ and $\theta_{02}$ (solid circles).

TOC Image

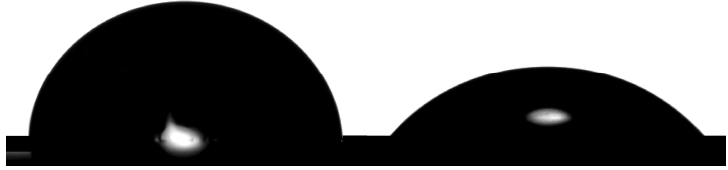